\documentstyle[12pt,psfig]{article}

\catcode`@=11
\long\def\@caption#1[#2]#3{\par\addcontentsline{\csname
  ext@#1\endcsname}{#1}{\protect\numberline{\csname
  the#1\endcsname}{\ignorespaces #2}}\begingroup
    \small
    \@parboxrestore
    \@makecaption{\csname fnum@#1\endcsname}{\ignorespaces #3}\par
  \endgroup}
\catcode`@=12

\input{epsf}
\ifx\epsffile\undefined
\def\epsffile#1#2#3#4]#5{}
\fi
\input{rotate}

\def\APP#1#2#3{Acta. Phys. Pol. {\bf B#1}, #2 (19#3)}

\def\MPLA#1#2#3{Mod. Phys. Lett. {\bf A#1}, #2 (19#3)}
\def\NPB#1#2#3{Nucl. Phys. {\bf B#1}, #2 (19#3)}

\def\PLB#1#2#3{Phys. Lett. {\bf B#1}, #2 (19#3)}
\def\PLBold#1#2#3{Phys. Lett. {\bf#1B}, #2 (19#3)}

\def\PRD#1#2#3{Phys. Rev. {\bf D#1}, #2 (19#3)}

\newcommand{\sign}{\:\!\rm{sign}\:\!}
\newcommand{\mgaugino}{M_{1/2}}
\newcommand{\mtop}{m_t}
\newcommand{\mweak}{M_{\rm{weak}}}
\newcommand{\mgut}{M_{\rm{GUT}}}

\newcommand{\tb}{\tan\beta}
\newcommand{\Pinput}{P_{\rm{input}}}

\newcommand{\ifb}{ \rm{ fb}^{-1}}

\newcommand{\gev}{\rm{ GeV}}
\newcommand{\tev}{\rm{ TeV}}

\newcommand{\gsim}{\lower.7ex\hbox{$\;\stackrel{\textstyle>}{\sim}\;$}}
\newcommand{\lsim}{\lower.7ex\hbox{$\;\stackrel{\textstyle<}{\sim}\;$}}
\begin{document}

\pagestyle{empty}
\begin{titlepage}
\def\thepage {}        

\title{\bf Multi-TeV Scalars are Natural \\
           in Minimal Supergravity \\ [1cm]}

\author{ {\small \bf Jonathan L. Feng$^a$,
Konstantin T. Matchev$^b$ and Takeo Moroi$^a$}\\
\\
{\small {\it ${}^a$School of Natural Sciences}}\\  
{\small {\it Institute for Advanced Study}}\\
{\small {\it Princeton, NJ 08540, USA}}\\ 
\\
{\small {\it ${}^b$Theoretical Physics Department}}\\  
{\small {\it Fermi National Accelerator Laboratory}}\\
{\small {\it Batavia, IL 60510, USA}}\\ }

\date{ }

\maketitle

   \vspace*{-12.5cm}

\noindent

\rightline{IASSNS-HEP-99-78}
\rightline{FERMILAB-PUB-99/226-T}

\vspace*{11.5cm}

\baselineskip=18pt

\begin{abstract}

{\normalsize For a top quark mass fixed
to its measured value, we find natural
regions of minimal supergravity parameter space where all squarks,
sleptons, and heavy Higgs scalars have masses far above 1 TeV and are
possibly beyond the reach of the Large Hadron Collider at CERN.  This
result is simply understood in terms of ``focus point''
renormalization group behavior and holds in any supergravity theory
with a universal scalar mass that is large relative to other
supersymmetry breaking parameters.  We highlight the importance of the
choice of fundamental parameters for this conclusion and for
naturalness discussions in general. }

\end{abstract}

\vfill
\end{titlepage}

\baselineskip=18pt
\pagestyle{plain}
\setcounter{page}{1}

The standard model with a fundamental Higgs boson suffers from a large
and unexplained hierarchy between the weak and Planck
scales~\cite{SM}.  Because supersymmetric theories are free of
quadratic divergences, however, this hierarchy is stabilized in
supersymmetric extensions of the standard model when the scale of
superpartner masses is roughly of order the weak scale
$\mweak$~\cite{SUSY}.  The promise of providing a natural solution to
the gauge hierarchy problem is the primary phenomenological motivation
for supersymmetry.

Because the requirement of naturalness places upper bounds on
superpartner masses, this criterion has important experimental
implications.  In a model-independent analysis, naturalness
constraints are weak for some superpartners, e.g., the squarks and
sleptons of the first two generations~\cite{nonuniv}.  However, in
widely studied scenarios where the scalar masses are unified at some
high scale, such as minimal supergravity, it is commonly assumed that
squark and slepton masses must all be $\lsim 1\ \tev$.  This bound
places all scalar superpartners within the reach of present and near
future colliders, and is a source of optimism in the search for
supersymmetry at the high energy and high precision frontiers.  We
show here, however, that this assumption is invalid, and in fact, it
is precisely in supergravity theories with a universal scalar mass
that {\em all\/} squark and slepton masses may naturally be far above
1 TeV.

Supersymmetric theories are considered natural if the weak scale is
not unusually sensitive to small variations in the fundamental
parameters.  Although the criterion of naturalness is inherently
subjective, its importance for supersymmetry has motivated several
groups to provide quantitative definitions of
naturalness~\cite{Ellis,Barbieri,Ross,deCarlos,Anderson,Strumia,%
Chankowski,Nath,nonsugra}.  In this analysis, we adopt the following
prescription:

\noindent (1) We consider the minimal supergravity framework with its
4+1 input parameters
\begin{equation}
\left\{ \Pinput \right\} = \left\{ m_0, \mgaugino, A_0, \tb,
\sign(\mu) \right\} \ ,
\end{equation}
where $m_0$, $\mgaugino$, and $A_0$ are the universal scalar mass,
gaugino mass, and trilinear coupling, respectively, $\tb = \langle
H_u^0 \rangle /\langle H_d^0 \rangle$ is the ratio of Higgs
expectation values, and $\mu$ is the Higgsino mass parameter.  The
first three parameters are at the grand unified theory (GUT) scale
$\mgut \simeq 2\times 10^{16}\ \gev$, i.e., the scale where the
U(1)$_Y$ and SU(2) coupling constants meet.

\noindent (2) The naturalness of each point ${\cal P} \in \{ \Pinput
\}$ is then calculated by first determining all the parameters of the
theory (Yukawa couplings, soft supersymmetry breaking masses, etc.),
consistent with low energy constraints. Renormalization group (RG)
equations are used to relate high and low energy boundary
conditions. In particular, at the weak scale, proper electroweak
symmetry breaking requires\footnote{The tree-level conditions are
displayed here for clarity of presentation.  In all numerical results
presented below, we use the full one-loop Higgs potential \cite{PBMZ},
minimized at the scale $m_0/2$, approximately where one-loop
corrections are smallest, as well as two-loop RG equations \cite{2loop
RGEs}, including all low-energy thresholds \cite{PBMZ,BMP}.}
\begin{eqnarray}
\frac{1}{2} m_Z^2 &=& \frac{m_{H_d}^2 - m_{H_u}^2 \tan^2\beta }
{\tan^2\beta -1} - \mu^2 \nonumber \\
&\equiv& f(m_{H_d}^2, m_{H_u}^2, \tan\beta) - \mu^2 \ , \label{mz} \\
2 B\mu &=& \sin 2 \beta \, (m_{H_d}^2 + m_{H_u}^2 + 2 \mu^2 ) \ ,
\end{eqnarray}
where $m_{H_u}^2$ and $m_{H_d}^2$ are the soft scalar Higgs masses,
and $B\mu$ is the bilinear scalar Higgs coupling.

\noindent (3) We choose to consider the following set of (GUT scale)
parameters to be free, independent, and fundamental:
\begin{equation}
\{ a_i \} = \{ m_0, \mgaugino, A_0, B_0, \mu_0 \} \ .
\label{fundamental}
\end{equation}

\noindent (4) All observables, including the $Z$ boson mass, are then
reinterpreted as functions of the fundamental parameters $a_i$, and
the sensitivity of the weak scale to small fractional variations in
these parameters is measured by the sensitivity
coefficients~\cite{Ellis,Barbieri}
\begin{equation}
c_i \equiv \left| \frac{\partial \ln m_Z^2}{\partial \ln a_i} \right|
\ .
\label{sensitivity}
\end{equation}

\noindent (5) Finally, we form the fine-tuning parameter
\begin{equation}
c = \max \{ c_i \} \ ,
\label{c}
\end{equation}
which is taken as a measure of the naturalness of point $\cal P$, with
large $c$ corresponding to large fine-tuning.

As is clear from the description above, several subjective choices
have been made, as they must be in any definition of naturalness.  The
choice of minimal supergravity in step (1), and particularly the
assumption of a universal scalar mass, plays a crucial role.
Deviations from this assumption will be considered below.

The choice of fundamental parameters in step (3) is also important and
varies throughout the literature.  An appealingly simple choice (see,
e.g., Ref.~\cite{Nath}) is $\{ a_i \} = \{ \mu \}$, where $\mu$ is to
be evaluated at the weak scale. This is equivalent to using $\mu^2$ as
a fine-tuning measure, since Eqs.~(\ref{mz}) and (\ref{sensitivity})
imply $c_{\mu} = 4\mu^2/m_Z^2$.  While generally adequate, this
definition is insensitive to large fine-tunings in the function $f$ of
Eq.~(\ref{mz}), as we will see below; such fine-tunings are accounted
for in the more sophisticated choice of Eq.~(\ref{fundamental}).

The top quark Yukawa $Y_t$ (sometimes along with other standard model
parameters, such as the strong coupling) is included among the
fundamental parameters in some studies~\cite{Ross,deCarlos,Anderson}
and not in others~\cite{Ellis,Strumia,Chankowski}.  This choice
typically attracts little comment, and attitudes toward it are at best
ambivalent~\cite{Barbieri}.  This ambiguity reflects, perhaps, a
diversity of prejudices concerning the fundamental theory of
flavor. It is important to note, however, that unlike the parameters
of Eq.~(\ref{fundamental}), $Y_t$ is not expected to be related to
supersymmetry breaking and is, in some sense, now measured, as it is
strongly correlated with the top quark mass $\mtop$.  For these
reasons, we find it reasonable to assume that in some more fundamental
theory, $Y_t$ is fixed to its measured value in a flavor sector
separate from the supersymmetry breaking sector, and we therefore do
not include it among the $a_i$. This choice is critical for our
conclusions, as will be discussed below.

In step (5), various other choices are also possible.  For example,
the $c_i$ may be combined linearly or in quadrature; we follow the
most popular convention.  In other prescriptions, the $c_i$ are
combined after first dividing them by some suitably defined average
$\overline{c}_i$ to remove artificial appearances of
fine-tuning~\cite{deCarlos,Anderson}.  We have not done this, but note
that such a normalization procedure typically reduces the fine-tuning
measure and would only strengthen our conclusions.

\begin{figure}[t!]
\centerline{\psfig{figure=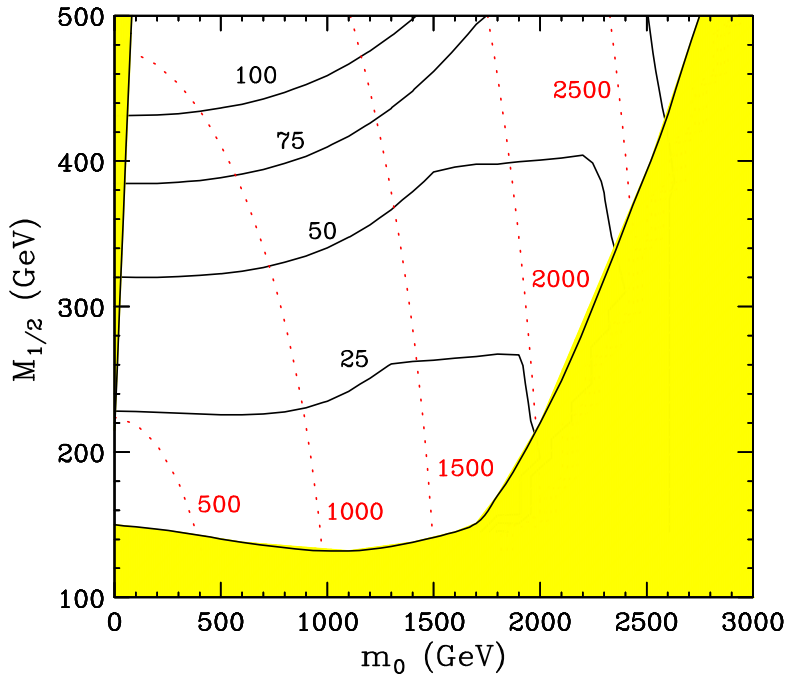,height=3.5in}}
\begin{center}
\parbox{5.5in}{
\caption[] {\small Contours of constant fine-tuning $c$ (solid) and
$m_{\tilde{u}_L}$ in GeV (dotted) in the $(m_0, \mgaugino)$ plane for
$\tb = 10$, $A_0=0$, and $\mu>0$. The shaded regions are excluded by
the requirement that the lightest supersymmetric particle be neutral
(top left) and by the chargino mass limit of 90 GeV (bottom and right).
\label{fig:finetune}}}
\end{center}
\end{figure}

Given the prescription for measuring naturalness described above, we
may now present our results.  In Fig.~\ref{fig:finetune}, contours of
constant $c$, along with squark mass contours, are presented for
$\tb=10$.  Moving from low to high $m_0$, the contours are determined
successively by $c_{\mu_0}$, $c_{M_{1/2}}$ and $c_{m_0}$.  The
naturalness requirement $c< 25$ ($c<50$) allows regions of parameter
space with $m_0 \approx 2\ \tev \ (2.4\ \tev)$.  More importantly,
regions with $m_0\gsim 2\ \tev$, where all squarks and sleptons have
masses well above 1 TeV, are as natural as the region with $(m_0,
\mgaugino) \lsim (1000\ \gev,400\ \gev)$, where squark masses
are below 1 TeV.

The naturalness of multi-TeV $m_0$, though perhaps surprising, may be
simply understood as a consequence of a ``focus point'' in the RG
behavior of $m_{H_u}^2$~\cite{focus}, which renders its value at
$\mweak$ highly insensitive to its value in the ultraviolet. Note that
for moderate and large $\tb$, Eq.~(\ref{mz}) implies that $m_Z^2$ is
insensitive to $m_{H_d}^2$ and is determined primarily by $m_{H_u}^2$.

Consider any set of minimal supergravity input parameters.  These
generate a particular set of RG trajectories, $m_i^2 |_{\rm{p}}(t),
M_i |_{\rm{p}}(t), A_i |_{\rm{p}}(t), \ldots$, where $t \equiv \ln
(Q/\mgut)$ and $Q$ is the renormalization scale.  Now consider another
set of boundary conditions that differs from the first by shifts in
the scalar masses.  The new scalar masses $m_i^2 = m_i^2 |_{\rm{p}}
+ \delta m_i^2$ satisfy the RG equations
\begin{equation}
\frac{d}{dt} m^2_i \sim \frac{1}{16\pi^2} \biggl[
- g^2 \mgaugino^2 + Y^2 A^2 + \sum_j Y^2 m_j^2 \biggr]
\label{rge}
\end{equation}
at one-loop, where positive numerical coefficients have been omitted,
and the sum is over all chiral fields $\phi_j$ interacting with
$\phi_i$ through the Yukawa coupling $Y$.  However, because the $m_i^2
|_{\rm{p}}$ are already a particular solution to these RG equations,
the deviations $\delta m_i^2$ obey the homogeneous equations
\begin{equation}
\frac{d}{dt} \delta m^2_i \sim \frac{1}{16\pi^2} \sum_j Y^2
\delta m_j^2 \ .
\label{schematic}
\end{equation}

Such equations are easily solved. Assume for the moment that the only
large Yukawa coupling is $Y_t$, i.e., $\tb$ is not extremely
large. Then $\delta m_{H_u}^2$ is determined from
\begin{equation}
\frac{d}{dt} \left[ \begin{array}{c} \delta m_{H_u}^2
\\ \delta m_{U_3}^2 \\ \delta m_{Q_3}^2 \end{array} \right]
= \frac{Y_t^2}{8\pi^2} \left[
\begin{array}{ccc}
3 & 3 & 3 \\
2 & 2 & 2 \\
1 & 1 & 1 \end{array} \right]
\left[ \begin{array}{c} \delta m_{H_u}^2
\\ \delta m_{U_3}^2 \\ \delta m_{Q_3}^2 \end{array} \right] \ ,
\end{equation}
where $Q_3$ and $U_3$ denote the third generation squark SU(2) doublet
and up-type singlet representations, respectively.  The solution
corresponding to the universal initial condition $\delta m_0^2 \,
(1,1,1)^T$ is

\begin {equation}
\! \left[ \! \! \begin{array}{c} \delta m_{H_u}^2
\\ \delta m_{U_3}^2 \\ \delta m_{Q_3}^2 \end{array} \! \! \right]
\! = \frac{\delta m_0^2}{2} \! \left\{
\left[ \! \begin{array}{c} 3 \\ 2 \\ 1 \end{array} \! \right]
{\rm exp} \! \left[ \int_0^t \frac{6Y_t^2}{8\pi^2} dt' \right]
-\left[ \! \! \begin{array}{r} 1 \\ 0 \\ -1 \end{array}
\! \! \right] \right\}.
\label{focus}
\end{equation}
For $t$ and $Y_t$ such that ${\rm exp}\! \left[ {6\over8\pi^2}
\int_0^t Y_t^2 dt' \right] = 1/3$, $\delta m_{H_u}^2 = 0$,
i.e., $m_{H_u}^2$ is independent of $\delta m_0^2$.
\begin{figure}[t!]
\centerline{\psfig{figure=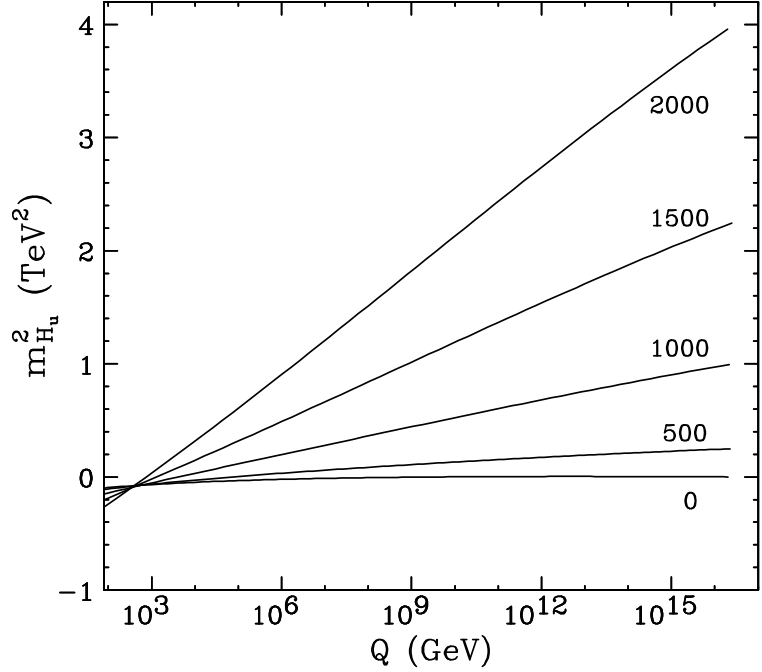,height=3.5in}}
\begin{center}
\parbox{5.5in}{
\caption[] {\small The RG evolution of $m_{H_u}^2$
for fixed $\mgaugino = 200$
GeV, $A_0=0$, $\tan\beta=10$, $\mu>0$, $\mtop = 175$ GeV, and several
values of $m_0$ (shown, in GeV).  The RG behavior of $m_{H_u}^2$
exhibits a focus point near the weak scale, where $m_{H_u}^2$ takes
its weak scale value $\sim -(300\ \gev)^2$, irrespective of $m_0$.
\label{fig:run}}}
\end{center}
\end{figure}

The RG evolution of $m_{H_u}^2$ in minimal supergravity is shown for
several values of $m_0$ in Fig.~\ref{fig:run}.  As expected, the RG
curves exhibit a focus (not a fixed) point, where $m_{H_u}^2$ is
independent of its ultraviolet value.  Remarkably, however, for the
physical top mass of $\mtop \approx 175\ \gev$, the focus point is very
near the weak scale.  Thus, the weak scale value of $m_{H_u}^2$ and,
with it, the fine-tuning parameter $c$ are highly insensitive to
$m_0$.  If the particular solution is natural (say, with all input
parameters near the weak scale), the new solution, even with very
large $m_0$, is also natural.

We have also checked numerically that the focusing effect persists
even for very large values of $\tan\beta$.  Indeed, in the limit $Y_t
= Y_b \gg Y_\tau$, Eq.~(\ref{schematic}) can be similarly solved
analytically, and one finds that focusing occurs for ${\rm
exp}\! \left[ {7\over8\pi^2} \int_0^t Y_t^2 dt' \right] = 2/9$.  For
the experimentally preferred range of top masses, the focus point is
again tantalizingly close to $\mweak$~\cite{inprep}.

The naturalness of multi-TeV $m_0$ has important implications for
collider searches.  Although $m_{H_u}^2$ is focused to the weak scale,
all other soft masses remain of order $m_0$.  {}From
Eqs.~(\ref{schematic}) and (\ref{focus}), we find that for $m_0 \gg
\mgaugino, A_0$, the physical masses of squarks, sleptons, and heavy
Higgs scalars are well-approximated by
\begin{eqnarray}
\tilde{t}_R : \sqrt{1/3}\, m_0 &\qquad&
\rm{All\ other\ } \tilde{q},\tilde{\ell} : m_0 \nonumber \\
\tilde{t}_L, \tilde{b}_L : \sqrt{2/3}\, m_0 &\qquad& \!
\quad H^{\pm}, A, H^0 : m_0 \ .
\end{eqnarray}
Exact values of $m_{\tilde{u}_L}$ are presented in
Fig.~\ref{fig:finetune}.  All squarks, sleptons, and heavy Higgs
scalars may therefore have masses $\gsim 1-2\ \tev$, and may be beyond
the reach of the Large Hadron Collider (LHC) and proposed linear
colliders.  The discovery of such heavy scalars then requires some
even more energetic facility, such as the envisioned muon or very
large hadron colliders.

\begin{figure}[t!]
\centerline{\psfig{figure=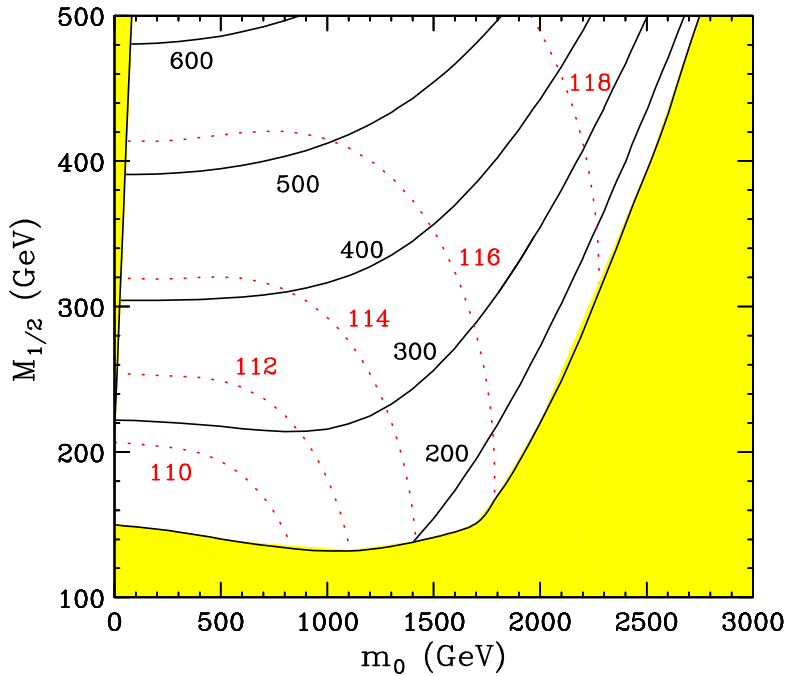,height=3.5in}}
\begin{center}
\parbox{5.5in}{
\caption[] {\small Contours of $\mu$ (solid)
and $m_h$ (dotted) in GeV for input
parameters as in Fig.~\ref{fig:finetune}.
\label{fig:mu}}}
\end{center}
\end{figure}

As may be seen from Fig.~\ref{fig:finetune}, however, fine-tuning
constraints do not allow multi-TeV $\mgaugino$.  A similar conclusion
applies to $\mu$, as may be seen in Fig.~\ref{fig:mu}. We therefore
expect all gauginos and Higgsinos to be within the kinematic reach of
the LHC. Note that some regions of low $\mu$ are unnatural. In these
regions, large cancellations in the function $f$ of Eq.~(\ref{mz})
occur, and the simple definition $c \propto \mu^2$ is inadequate.

In addition to the gauginos and Higgsinos, the lightest Higgs boson
is, of course, still required to be light.  Contours of lightest Higgs
mass $m_h$ are also presented in Fig.~\ref{fig:mu}.  Very heavy top
and bottom squarks increase $m_h$ through radiative corrections: for
low $\mgaugino$, $m_h$ increases by roughly $6$ GeV as $m_0$ increases
from 500 GeV to 2 TeV.  However, in the multi-TeV $m_0$ scenario,
naturalness requires $A_0 \sim \mweak$ (see below), and so left-right
squark mixing is suppressed. The upper bound on $m_h$ in
Fig.~\ref{fig:mu} is thus approximately $120\ \gev$, well below limits
achieved for TeV squarks with maximal left-right mixing, and within
the 3-5$\sigma$ discovery range of Higgs searches at the Tevatron with
luminosity $10-30\ \ifb$ \cite{Higgs report}.

The focus point analysis presented above (for small $Y_b$) relied
heavily on the universality of the $H_u$, $U_3$ and $Q_3$ soft masses.
It is not hard to show, however, that GUT scale boundary conditions of
the form $(m_{H_u}^2, m_{U_3}^2, m_{Q_3}^2) = (1,1-x,1+x)$, for any
$x$, also exhibit the focus point behavior.  With respect to the other
supersymmetry breaking parameters, the focus point is fairly robust.
The mechanism is independent of all other scalar masses. Also, in the
analysis above, {\em any} natural particular solution would do.
Arbitrary and non-universal gaugino masses and trilinear couplings of
order $\mweak$ are therefore allowed. (Similarly, deviations in
$m_{H_u}^2$, $m_{U_3}^2$, and $m_{Q_3}^2$ of order $\mweak^2$ do not
destabilize the focus point.)  Note, however, that multi-TeV gaugino
masses and $A$ parameters are not allowed.  The required hierarchy
between the scalar masses and the gaugino mass, $A$, and $\mu$
parameters may result from an approximate U(1)$_{R+PQ}$ symmetry or
from the absence of singlet $F$ terms~\cite{Rsymm}. $B\mu$ may also be
suppressed by such a symmetry, and so leads to an experimentally
viable scenario with naturally large $\tb\approx m_{H_d}^2/(B\mu)$,
which is typically difficult to realize~\cite{largetb}.

Although the focus point mechanism depends on a relation between
$\mtop$ and $\ln ({M_{\rm GUT}\over M_{\rm weak}})$,
it is not extraordinarily
sensitive to these values.  The focus point is still near the weak
scale if $\mtop$ is varied within its experimental uncertainty of 5
GeV, and, in fact, natural regions with multi-TeV $m_0$ are also
possible if the high scale is raised to $\sim 10^{18}
\ \gev$~\cite{inprep}.

We stress, however, that if $Y_t$ is included among the free and
fundamental parameters, multi-TeV $m_0$ would be considered
unnatural. For example, for $\tb=10$ and $A_0=0$, $c_{Y_t} < 25$ (50)
corresponds to $m_0 \lsim 500\ \gev$ (800 GeV)~\cite{inprep}.  We have
presented above our rationale for not including $Y_t$ among the $a_i$,
although a definitive resolution of this issue most likely requires an
understanding of the fundamental theory of flavor.

In conclusion, for moderate and large $\tb$, multi-TeV scalars are
natural in minimal supergravity.  In view of this result, the
discovery of squarks, sleptons, and heavy Higgs scalars may be
extremely challenging even at the LHC.  In addition, it is not
surprising that these scalars have so far escaped detection, as
present bounds are far from excluding most of the natural parameter
space.  Finally, it is tempting to speculate that what appears to be
an accidental conspiracy between $\mtop$ and the ratio of high to weak
scales may find some fundamental explanation.  If gauginos and
Higgsinos are discovered, but all supersymmetric scalars escape
detection at the LHC, the preservation of the naturalness motivation
for supersymmetry, as currently understood, will require either an
explanation of large cancellations between supersymmetry breaking soft
masses at the weak scale, or the above scenario with a top mass fixed
to be near 175 GeV. The latter possibility is, in our view, far more
compelling and is supported by experimental data.

{\em Acknowledgments} --- We are grateful to K.~Agashe, M.~Drees, and
L.~Hall for correspondence and conversations, and to the Aspen Center
for Physics for hospitality.  This work was supported in part by DOE
under contracts DE--FG02--90ER40542 and DE--AC02--76CH03000, by the
NSF under grant PHY--9513835, through the generosity of Frank and
Peggy Taplin (JLF), and by a Marvin L.~Goldberger Membership (TM).

\newpage

\vfill

\begin{thebibliography}{99}
\frenchspacing

\bibitem{SM}
K.~Wilson, unpublished;\\
L.~Susskind, \PRD{20}{2619}{79};\\
G.~'t~Hooft, in {\em Recent Developments in Gauge Theories},
ed.~G.~'t~Hooft {\em et al.}, (Plenum Press, New York, 1980), p.~135.

\bibitem{SUSY}
L.~Maiani, in {\em Proc. Gif-sur-Yvette Summer School} (Paris, 1980),
p.~3;\\
E.~Witten, \NPB{188}{513}{81};\\
M.~Veltman, \APP{12}{437}{81};\\
R.~Kaul, \PLBold{109}{19}{82}.

\bibitem{nonuniv}
S.~Dimopoulos and G.~F.~Giudice,
\PLB{357}{573}{95}, {\tt hep-ph/9507282};\\
A.~Pomarol and D.~Tommasini,
\NPB{466}{3}{96}, {\tt hep-ph/9507462}.

\bibitem{Ellis}
J.~Ellis, K.~Enqvist, D.~V.~Nanopoulos, and F.~Zwirner,
\MPLA{1}{57}{86}.

\bibitem{Barbieri}
R.~Barbieri and G.~F.~Giudice, \NPB{306}{63}{88}.

\bibitem{Ross}
G.~G.~Ross and R.~G.~Roberts,
\NPB{377}{571}{92}.

\bibitem{deCarlos}
B.~de~Carlos and J.~A.~Casas,
\PLB{309}{320}{93}, {\tt hep-ph/9303291}.

\bibitem{Anderson}
G.~W.~Anderson and D.~J.~Castano,
\PLB{347}{300}{95}, {\tt hep-ph/9409419};\\
\PRD{52}{1693}{95}, {\tt hep-ph/9412322};\\
\PRD{53}{2403}{96}, {\tt hep-ph/9509212}.

\bibitem{Strumia}
P.~Ciafaloni and A.~Strumia,
\NPB{494}{41}{97}, {\tt hep-ph/9611204};\\
G.~Bhattacharyya and A.~Romanino,
\PRD{55}{7015}{97}, {\tt hep-ph/9611243}; \\
R.~Barbieri and A.~Strumia,
\PLB{433}{63}{98}, {\tt hep-ph/9801353}; \\
L.~Giusti, A.~Romanino, and A.~Strumia,
\NPB{550}{3}{99}, {\tt hep-ph/9811386}.

\bibitem{Chankowski}
P.~H.~Chankowski, J.~Ellis, and S.~Pokorski,
\PLB{423}{327}{98}, {\tt hep-ph/9712234};\\ 
P.~H.~Chankowski, J.~Ellis, M.~Olechowski, and S.~Pokorski,
\NPB{544}{39}{99}, {\tt hep-ph/9808275}.

\bibitem{Nath}
K.~L.~Chan, U.~Chattopadhyay, and P.~Nath,
\PRD{58}{096004}{98}, {\tt hep-ph/9710473}.

\bibitem{nonsugra}
D.~Wright, preprint UW/PT-97/27, {\tt hep-ph/9801449}; \\
G.~L.~Kane and S.~F.~King, \PLB{451}{113}{99},
{\tt hep-ph/9810374}.

\bibitem{PBMZ}
D.~Pierce, J.~Bagger, K.~Matchev, and R.-J.~Zhang,
\NPB{491}{3}{97}, {\tt hep-ph/9606211}.

\bibitem{2loop RGEs}
S.~Martin and M.~Vaughn,
\PRD{50}{2282}{94}, {\tt hep-ph/9311340}.

\bibitem{BMP}
J.~Bagger, K.~Matchev, and D.~Pierce,
\PLB{348}{443}{95}, {\tt hep-ph/9501277}.

\bibitem{focus}
J.~L.~Feng and T.~Moroi, 
preprint IASSNS-HEP-99-65, {\tt hep-ph/9907319}.

\bibitem{inprep}
J.~L.~Feng, K.~T.~Matchev, and T.~Moroi, in preparation.

\bibitem{Higgs report}
Final Report of the Higgs Working Group,
http://fnth37.fnal.gov/higgs/draft.html.

\bibitem{Rsymm}
See, e.g.,
J.~L.~Feng, N.~Polonsky, and S.~Thomas, 
\PLB{370}{95}{96}, {\tt hep-ph/9511324}; \\
J.~L.~Feng, C.~Kolda, and N.~Polonsky,
\NPB{546}{3}{99}, {\tt hep-ph/9810500}.

\bibitem{largetb}
A.~E.~Nelson and L.~Randall, \PLB{316}{516}{93},
{\tt hep-ph/9308277}.

\end{thebibliography}
\end{document}